\documentclass[12pt, letterpaper,fleqn]{article} 
\usepackage[onehalfspacing]{setspace}
\usepackage [english,american]{babel}  
				
\usepackage[square]{natbib}	
\usepackage[margin=1in]{geometry}
\usepackage[mathscr]{euscript}
\usepackage{amssymb}
\usepackage{amsthm}
\usepackage{amsmath}
\usepackage{amsfonts}
\usepackage{csquotes}
\usepackage{epstopdf}
\usepackage{graphicx}
\usepackage{epigraph}
\usepackage{stackengine}
\AtBeginEnvironment{quote}{\small}
\usepackage{multirow,array}
\usepackage{booktabs}
\usepackage{bm}

\def\app#1#2{%
  \mathrel{%
    \setbox0=\hbox{$#1\sim$}%
    \setbox2=\hbox{%
      \rlap{\hbox{$#1\propto$}}%
      \lower1.1\ht0\box0%
    }%
    \raise0.25\ht2\box2%
  }%
}
\def\approxprop{\mathpalette\app\relax}

\newcommand*\vect[1]{\mathbf{\bm{#1}}}

\usepackage{xpatch}
\usepackage{hyperref}
\hypersetup{colorlinks=true,citecolor=blue}

\title{Bayesian Inference and the Principle of Maximum Entropy}
\author{Duncan K. Foley\footnote{Department of Economics, The New School for Social Research. Email: foleyd@newschool.edu} and Ellis Scharfenaker\footnote{Department of Economics, University of Utah. Email: ellis.scharfenaker@economics.utah.edu}}

\date{\today}                                           

\begin{document}


\maketitle

\begin{abstract}
\singlespacing
Bayes' theorem incorporates distinct types of information through the likelihood and prior. Direct observations of state variables enter the likelihood and modify posterior probabilities through consistent updating. Information in terms of expected values of state variables modify posterior probabilities by constraining prior probabilities to be consistent with the information. Constraints on the prior can be exact, limiting hypothetical frequency distributions to only those that satisfy the constraints, or be approximate, allowing residual deviations from the exact constraint to some degree of tolerance. When the model parameters and constraint tolerances are known, posterior probability follows directly from Bayes' theorem. When parameters and tolerances are unknown a prior for them must be specified. When the system is close to statistical equilibrium the computation of posterior probabilities is simplified due to the concentration of the prior on the maximum entropy hypothesis. The relationship between maximum entropy reasoning and Bayes' theorem from this point of view is that maximum entropy reasoning is a special case of Bayesian inference with a constrained entropy-favoring prior.
\end{abstract}


\newpage
\section{Two forms of statistical information}

Information structures statistical analysis in two distinct ways.\footnote{See \citet[ch 4-6]{FoleyScharfenaker2023} manuscript for details.} We can illustrate these two forms of information in the context of the multinomial model of an urn of unknown composition represented by a normalized frequency distribution. In this case the unknown variable is the frequency distribution describing the composition of the urn. Each hypothesized composition frequency distribution determines the likelihood of drawing any specific sample. Bayesian inference constructs a posterior distribution over the space of hypothetical frequency distributions conditional on sample data as the product of a prior probability that assigns a probability to each hypothetical frequency distribution and the likelihood that incorporates the sample data.

One form of information, then, is a sample comprising observations from the system. Observational information shapes posterior inferences through the Bayesian likelihood. Drawing and observing a red ball from an urn of unknown composition, for example, requires consistent updating of posterior probabilities over hypothetical compositions of the urn to assign a zero probability to any hypothetical composition with zero red balls. Information in the form of a sample of direct observations is common in the social sciences where researchers have limited direct access to quantitative and qualitative reports of individual phenomena. 

A second form of information is theoretically or experimentally determined expected values of relevant quantities that constrain the hypothetical frequency distributions describing the composition of the urn. Information in the form of experimental or theoretical constraints shapes posterior inference through the Bayesian prior by assigning zero or low prior probability to hypothesized distributions that fail to meet the constraints. These constraints may also be expressed in terms of conditional expectations, or any other limits on the hypothetical frequency distribution describing the system state. Information in the form of expectations is common in physical sciences where direct observation of, e.g. the momentum or position of individual molecules in a thermodynamic system is not feasible, but macroscopic experimental averages such as the energy or temperature of a gas can be measured or predicted. 

A general framework for drawing consistent posterior degrees of belief must be able to incorporate both forms of information as a mixture of moment constraints and direct sample observations.\footnote{As \citet{Jaynes1988} acknowledged, ``In almost all real problems of scientific inference we need to take into account our total state of knowledge, only part (or sometimes none) of which consists of frequencies."}

\citet{Jaynes1983} demonstrated that information in the form of expectations that constrain hypothesized frequencies describing a system such as the composition of an urn can be incorporated into statistical estimation by maximizing the Shannon informational entropy of the hypothesized distribution subject to whatever constraints apply.\footnote{See also \citet{Soofi1994,Soofi2000} for a brief overview on information theoretic methods.} More recently, \citet{Golan2018} has demonstrated for a broad class of model settings why using hypothesized frequency distributions that maximize informational entropy subject to constraints as estimates of the system state lead to robust and convincing statistical conclusions. The logic behind the principle of maximum entropy is that maximizing informational entropy adds the least possible information to the explicit information represented by the constraints and therefore is the procedure least likely to bias estimates. 

\citet{Jaynes1957} also demonstrated that the principle of maximum entropy inference is closely related to the concept of statistical equilibrium from statistical physics, which turns out to have important implications for the connection between the principle of maximum entropy and Bayesian inference.

While Jaynes strongly advocated and advanced both the principle of maximum entropy \citep{Jaynes1979} and Bayes' theorem as logical methods of inference \citep{Jaynes2003} his account of the relationship between the two was less than complete.

\begin{quote}We have at present two principles, Bayes' theorem and MAXENT, that are held to have some fundamental status. The practitioners of the art sometimes use one, sometimes the other; but beginners and critics alike seems puzzled by how we choose between them. How are these principles related to each other? Are they mutually consistent or in conflict? What is the proper place of each in our toolbox?\citep{Jaynes1988a}
\end{quote}

Since 1981 the MAXENT workshop, organized in part by Jaynes, has been exploring the relationship between maximum entropy and Bayesian methods in physical and social sciences. These workshops have significantly advanced our understanding of these two methods, but there remains some mystery and disagreement about the relationship between the principle of maximum entropy and Bayes' theorem, particularly on the question of which is more fundamental to statistical inference. In some cases \citep{Caticha2007a,Caticha2006, Williams1980, Zellner1988} Bayes' theorem appears as a special case of the principle of maximum entropy, while in others \citep{Seidenfeld1986, Campencourt1981, Skyrms1985, Good1992} Bayes' theorem appears as the fundamental principle of inference with maximum entropy arising as a special case.   

While \citet{Jaynes1988} recognized that a logical method of inference should incorporate both information in the form of direct observations of the state space that generate a Bayesian likelihood as well as theoretical or experimental information that constrains hypothesized state space configurations, he often contrasts these types of information in separate settings. Rather than demonstrating how both types of information influence the inferred posterior distribution by shaping the prior and likelihood, he obscures the problem setting by focusing on the ways frequentist methods such as the Darwin-Fowler method \citep{DarwinFolwer1922}\footnote{This method derives the distributional expression of the partition of energy in an ensemble of a large number of microsystems with a given total energy.}, are less convincing ways of arriving at the same result as one would get by maximizing the information entropy subject to constraints. 

\begin{quote}
If a statement $d$ referring to a probability distribution in space $S$ is testable (for example, if it specifies a mean value $\langle f \rangle$ for some function $f(i)$ defined on $S$), then it can be used as a constraint in PME; but it cannot be used as a conditioning statement in Bayes' theorem because it is not a statement about any event in $S$ or any other space. Conversely, a statement $D$ about an event in the space $S^n$ (for example, an observed frequency) can be used as a conditioning statement in applying Bayes' theorem, whereupon it yields a posterior distribution on $S^n$ which may be contracted to a marginal distribution on $S$; but $D$ cannot be used as a constraint in applying PME in space $S$, or about any probability distribution over $S$... whether we use maximum entropy in space $S$ with a constraint fixing an average $\langle f \rangle$ over a \underline{probability distribution}, or apply Bayes' theorem in $S^n$ with a conditioning statement fixing a numerically equal average $\bar{f}$ over \underline{sample values}, we obtain for large $n$ identical distributions in the space $S$... PME leads us directly to the same final result, without any need to go into a higher space $S^n$ and carry out passage to the limit $n\rightarrow \infty$ by saddle-point integration.\citep{Jaynes1979}
\end{quote}

The ways the principle of maximum entropy and Bayes' theorem have been contrasted and related are uneven have led to considerable confusion. In hopes of clarifying how we ``take into account our total state of knowledge", we develop the argument here that Bayes' theorem is the logical starting point for incorporating and updating any type of information. When information is in the form of direct observations of state variables it modifies posterior probabilities through the likelihood. Theoretical and experimental information on expectations or other properties of hypothesized frequency distributions modifies posterior probabilities through a constrained entropy prior that assigns higher prior probability to hypotheses that have higher entropy and meet the constraints. 

The principle of maximum entropy is a logically compelling way of incorporating experimental and theoretical information into the prior. When we use the principle of maximum entropy to derive estimates of frequency distributions describing the system state, we encounter two possibilities. When we know the exact or approximate values of the constraints (as with experimental information) the Lagrange multipliers associated with the constraint are known and we condition all unknown model parameters on this information. When we do not know the values of the constraints (as with theoretical determined expectations whose experimental value is unknown) we must assign a prior over those values and include our uncertainty about the Lagrange multipliers, now model parameters, to assign a posterior distribution to the space of hypotheses. When the system is close to statistical equilibrium the prior over hypotheses will be sharply peaked at the entropy maximizing hypothetical distribution and the posterior distribution will concentrate on the entropy maximizing hypothetical distribution.  

\section{Systems, State Spaces, and Data}

A system about which we want to make inferences is defined by a state space, which we denote $\mathcal{X}$, with individual states $x\in \mathcal{X}$. We define the state space as the product of $R$ subspaces $\mathcal{X} = \mathcal{X}_1 \times \ldots \times \mathcal{X}_R$, so that a state $x=\{x_1,\ldots,x_R\}$. A Bayesian hypothesis for the system is a normalized assignment of frequency weights $q[.]:\mathcal{X}\rightarrow [0,1],\sum_\mathcal{X}q[x]=1$. When the state space $\mathcal{X}$ is a finite set with $K$ elements we use the notational convention that $q[\mathcal{X}]$ is a $K$-dimensional vector representing the \textit{hypothesis distribution} over states. 

For $N$ exchangeable data observations from a system with $K$ states we represent these data with the sufficient statistic describing the macrostate of the system $\vect{n}=\{n_1,\ldots,n_K\}$ describing the frequency of observations for each $K$ categories. We can also express the data in terms of the relative frequencies of each state, which we refer to as the \textit{data distribution} $p[\mathcal{X}]=\{p_1=\frac{n_1}{N},\ldots,p_K=\frac{n_K}{N}\}$. Under exchangeability any data with the same $\vect{n}$ is assigned an equal probability and the likelihood for a sample $\vect{n}$ conditional on a hypothesis of relative frequencies $q[\mathcal{X}]=\{q_1,\ldots,q_K\}$ is:

\begin{align}
\label{eq:multlike}
\mathcal{P}[p[\mathcal{X}]|q[\mathcal{X}]]
\propto q_1^{n_1} q_2^{n_2} \ldots q_{K}^{n_{K}} = e^{N p[\mathcal{X}] \cdot \log[q[\mathcal{X}]]}\propto e^{-N H[p[\mathcal{X}]\|q[\mathcal{X}]]}
\end{align}

\noindent where $H[p[\mathcal{X}]\|q[\mathcal{X}]]$ is the relative entropy (or Kullback-Leibler divergence) of the hypothesis distribution $q[\mathcal{X}]$ to the data distribution $p[\mathcal{X}]$. Because we can express any state space in terms of a multinomial system through an appropriate coarse-graining, the Multinomial likelihood acts as a type of ``universal” likelihood that is an appropriate setting for illustrating the fundamental relationship between Bayes' theorem and the principle of maximum entropy \citep{Kullback1967}.

Posterior inference requires the addition of a prior over the hypotheses $q[\mathcal{X}]$. If, following \citet{Jaynes1983} and \citet[Ch. 4]{Golan2018}, we adopt prior probability that favors hypothesized frequency distributions that have a higher Shannon informational entropy, we minimize the information the prior imposes on the analysis of the system. A Shannon entropy favoring prior is expressed as:

\begin{align}
\label{eq:entprior}
\mathcal{P}[q[\mathcal{X}]]\propto e^{H[q[\mathcal{X}]]}=e^{-q[\mathcal{X}]\cdot \log[q[\mathcal{X}]]}
\end{align}

The posterior distribution of the hypothetical frequencies $q[\mathcal{X}]$ conditional on a data distribution $p[\mathcal{X}]$ with an entropy favoring prior is a fundamental statistical model for drawing inferences from multinomial systems and takes the form\footnote{See \citet{Soofi1995, Soofi2002} for details of estimation of this class of models}:

\begin{align}
\label{eq:post}
\mathcal{P}[q[\mathcal{X}]|p[\mathcal{X}]]\propto e^{H[q[\mathcal{X}]]}e^{-N H[p[\mathcal{X}]\|q[\mathcal{X}]]}
\end{align}

\section{Model Constraints}

Any information we have about a system constrains the class of models we consider in assigning posterior probabilities. Incorporating new information into probability assignments, either in the form of observed data that enter the likelihood or in the form of experimental or theoretical considerations that constrain the prior, modifies the joint data-hypothesis space as expressed in the posterior distribution. Bayes' theorem consistently reallocates probability from the class of models inconsistent with the information to the class of models that remain feasible. These considerations imply that  when we have information in the form of a mixture of direct observations that partially reveal the state of the system as well as theoretical or experimental information, we should adopt a constrained entropy-favoring prior to reflect the degree to which the hypothesis meets the constraints and then apply the observations through a likelihood function to the prior in order to compute the posterior conditional on the available information.

Given a well-defined state space \(\mathcal{X}\), we define an approximate model constraint on a hypothesis \(q[\mathcal{X}]\) based on a \(D\)-dimensional residual function \(\vect{g}_{\vect{\theta }}[\cdot]:\mathcal{X}\rightarrow \mathcal{R}^D\) such that $\vect{g}_{\vect{\theta }}[\mathcal{X}]$ is a $D$-dimensional vector of residual differences of the system from the model predictions described by a vector of parameters $\vect{\theta}$.  The constraint on the hypotheses generated by this system of residuals can be written in terms of the expected squared residuals: \(|\vect{g}_\vect{\theta}[\mathcal{X}]|^2\cdot q[\mathcal{X}] = \vect{s}^2\). While it is possible to impose the constraints exactly, exact constraints effectively overdetermine the system. Common statistical practice avoids this problem by permitting the constraints to hold only to some degree of approximation, which we represented by a vector of tolerances $\vect{s}^2$. These tolerance parameters increases the number of degrees of freedom of the system, effectively transforming the overdetermined problem into an underdetermined problem. 

The Constrained Maximum Entropy (CME) problem with approximate satisfaction of the residual constraints that defines the prior over hypothesis space $q[\mathcal{X}]$ is:

\begin{align}
&\max_{q[\mathcal{X}]\in \mathcal{Q}_K} -\sum_\mathcal{X}q[\vect{x}]\log[q[\vect{x}]]\\
&\text{subject to } \sum_\mathcal{X}|\vect{g}_\vect{\theta}[\vect{x}]|^2q[\vect{x}]= |\vect{g}_\vect{\theta}[\mathcal{X}]|^2\cdot q[\mathcal{X}] = \vect{s}^2 
\end{align}

\noindent where the normalized CME prior distribution lies in the $K$-dimensional unit simplex $\mathcal{Q}_K = \{q[\mathcal{X}]|\vect{1}_K\cdot q[\mathcal{X}] = 1\}$

This problem has the Lagrangian and first-order conditions, which are necessary and sufficient to characterize the solution:

\begin{align}
\label{eq;dualconstmaxentlag}
\mathcal{L}[q[\mathcal{X}],\vect{\lambda}]&= -\sum_\mathcal{X}q[\vect{x}]\log[q[\vect{x}]] - \vect{\lambda}\cdot \left(\sum_\mathcal{X}|\vect{g}_\vect{\theta}[\vect{x}]|^2q[\vect{x}] -\vect{s}^2\right)\notag \\
\frac{\partial \mathcal{L}}{\partial q[\vect{x}]} &= -\log[q[\vect{x}]] - \vect{\lambda}\cdot |\vect{g}_\vect{\theta}[\vect{x}]|^2 = 0\notag\\
\log[q[\vect{x}]] &= - \vect{\lambda}\cdot |\vect{g}_\vect{\theta}[\vect{x}]|^2\notag\\
\hat{q}_{\vect{\theta},\vect{\lambda}}[\vect{x}] &= \frac{e^{- \vect{\lambda}\cdot |\vect{g}_\vect{\theta}[\vect{x}]|^2}}{\sum_\mathcal{X} e^{- \vect{\lambda}\cdot |\vect{g}_\vect{\theta}[\vect{x}]|^2}}
\end{align}

Where $\vect{\lambda}$ is an $D$-dimensional vector of Lagrange multipliers corresponding to the residuals. Writing the Lagrange multipliers as inverse variances, $\vect{\lambda} = \frac{1}{\vect{\sigma}^2}$, we can see that the constrained maximum entropy distribution with approximate constraints on the mean squared residual deviations of the system is a joint-normal distribution. Maximizing the entropy subject to the approximate constraints leads to a reference CME distribution that is parameterized by the Lagrange multipliers on the constraints. This expression is the Legendre transformation that is widely used in statistical mechanics.

\begin{align}
\label{eq:dualconstmaxentnorm}
\hat{q}_{\vect{\theta},\vect{\sigma}}[\vect{x}]
= \frac{e^{- \vect{1}_D \cdot \left|\frac{\vect{g}_\vect{\theta}[\vect{x}]}{\vect{\sigma}}\right|^2}}{\sum_\mathcal{X} e^{- \vect{1}_D \cdot \left|\frac{\vect{g}_\vect{\theta}[\vect{x}]}{\vect{\sigma}}\right|^2}}
\end{align}

\noindent where $\vect{1}_D$ is a $D$-dimensional vector of ones. If we know the constraint parameters \(\pmb{\theta }\) and the tightness of the constraints \(\pmb{\sigma }\), the posterior conditional on this information and the data is:

\begin{align}
\label{eq:posterior}
\mathcal{P}[q[\mathcal{X}]|p[\mathcal{X}], \pmb{\theta },\pmb{\sigma }]\propto e^{H[q[\mathcal{X}]]-\pmb{1}_D\cdot \left| \frac{\vect{g}_{\vect{\theta }}[\mathcal{X}]}{\pmb{\sigma }}\right| {}^2\cdot q[\mathcal{X}]}e^{-N H[p[\mathcal{X}]\|q[\mathcal{X}]]}
\end{align}

\noindent and posterior inference is a straightforward application of Bayes' theorem. If we do not know \(\pmb{\theta },\pmb{\sigma}\), we have to supply a prior over these parameters in order to derive a posterior joint probability over the hypothesis, \(\pmb{\theta },\pmb{\sigma },q[\mathcal{X}]\),
conditional on the data:

\begin{align}
\mathcal{P}[\pmb{\theta },\pmb{\sigma },q[\mathcal{X}]|p[\mathcal{X}]]&\propto \mathcal{P}[\pmb{\theta },\pmb{\sigma }]\mathcal{P}[q[\mathcal{X}]|p[\mathcal{X}],\pmb{\theta },\pmb{\sigma }]\\
\label{eq:contentprior1}&= \mathcal{P}[\pmb{\theta },\pmb{\sigma }]e^{H[q[\mathcal{X}]]-\pmb{1}_D\cdot \left| \frac{\vect{g}_{\vect{\theta }}[\mathcal{X}]}{\pmb{\sigma }}\right| {}^2\cdot q[\mathcal{X}]}e^{-N H[p[\mathcal{X}]\|q[\mathcal{X}]]} \\
\label{eq:contentprior2}&=\mathcal{P}[\pmb{\theta },\pmb{\sigma }]e^{-H\left[q[\mathcal{X}]\left\|\hat{q}_{\pmb{\theta},\pmb{\sigma }}[\mathcal{X}]\right.\right]}e^{-N H[p[\mathcal{X}]\|q[\mathcal{X}]]}
\end{align}

The equation above demonstrates that there are two mathematically equivalent ways to formulate the constrained entropy-favoring prior. Eq~\ref{eq:contentprior1} expresses the prior as tradeoff between the entropy of the hypothesis distribution and the degree to which it approximately satisfies the constraints. Eq~\ref{eq:contentprior2} indirectly expresses the same concept as the divergence of the constrained maximum entropy distribution, which is the best one can do in trading off entropy and the constraint, from the hypothesis distribution. This second formulation of the constrained entropy-favoring prior emphasizes the role of the CME distribution as a benchmark against which alternative hypotheses have to compete.

If we want to draw inferences over the model parameters and degrees of approximation, \(\pmb{\theta },\pmb{\sigma }\), then the hypothetical frequencies \(q[\mathcal{X}]\) are {``}nuisance{''} parameters that we want to marginalize over:

\begin{align}
\mathcal{P}[\pmb{\theta },\pmb{\sigma }|p[\mathcal{X}]]&\propto \mathcal{P}[\pmb{\theta},\pmb{\sigma }]\int_\mathcal{Q} e^{H[q[\mathcal{X}]]-\pmb{1}_D\cdot \left| \frac{\vect{g}_{\vect{\theta }}[\mathcal{X}]}{\pmb{\sigma }}\right| {}^2\cdot q[\mathcal{X}]}e^{-N
H[p[\mathcal{X}]\|q[\mathcal{X}]]}dq[\mathcal{X}]
\end{align}

This integral is not in general evaluable except by numerical approximation for specific data. If, however, the term \(e^{H[q[\mathcal{X}]]-\pmb{1}_D\cdot \left| \frac{\vect{g}_{\vect{\theta }}[\mathcal{X}]}{\pmb{\sigma }}\right| {}^2\cdot q[\mathcal{X}]}\) is sharply
peaked at its maximum, \(\hat{q}_{\pmb{\theta },\pmb{\sigma }}[\mathcal{X}]\), or, equivalently, the system is close to statistical equilibrium given the constraints, then almost all the weight of the integral will be on the CME distribution, and we can use saddle-point integration for approximating the integral.\footnote{Saddle-point integration approximates integrals of the form, $\int e^{N g[x]}dx$. When the integral is sharply peaked it can be approximated by the maximum value of the integrand, obtained at the point $\hat{x}$ that maximizes the exponent of $g[x]$. Expanding the exponent around the maximum: \[g[x] \approx g[\hat{x}]-\frac{1}{2}\left |g''[\hat{x}]\right |(x-\hat{x})^2+\cdots\] At the maximum, the first derivative $g'[\hat{x}]=0$ and $g''[\hat{x}]<0$. When the integrand is negligibly small outside the neighborhood of $\hat{x}$ the integral is approximated with the standard Gaussian integral: \[\int e^{-N g[x]}dx\approx \sqrt{\frac{2\pi}{N\left |g''[\hat{x}]\right |}}e^{-N g[\hat{x}]}\propto e^{-N g[\hat{x}]}\] Extremizing the integrand of partition functions are often used in statistical physics, for example to find the standard formula for pressure in the Gibbs canonical ensemble or in deriving Stirling{'}s approximation for \(N!\). } In our case we want to approximate the partition function of \ref{eq:posterior}:

\begin{align}
\label{eq:sp-int}
\mathcal{Z}[p[\mathcal{X}], \vect{\theta} ,\vect{\sigma} ]&=\int_\mathcal{Q} e^{H[q[\mathcal{X}]]-\pmb{1}_D\cdot \left| \frac{\vect{g}_{\vect{\theta }}[\mathcal{X}]}{\pmb{\sigma}}\right| {}^2\cdot q[\mathcal{X}]}e^{-N H[p[\mathcal{X}]\|q[\mathcal{X}]]}dq[\mathcal{X}] \\
& \approx \frac{1}{\mathcal{Z}[\hat{q}_{\pmb{\theta },\pmb{\sigma }}[\mathcal{X}]]} e^{-N H\left[p[\mathcal{X}]\left\|\hat{q}_{\pmb{\theta },\pmb{\sigma }}[\mathcal{X}]\right.\right]}\\
& \propto e^{-N H\left[p[\mathcal{X}]\left\|\hat{q}_{\pmb{\theta },\pmb{\sigma }}[\mathcal{X}]\right.\right]}
\end{align}

\noindent where $\mathcal{Z}[\hat{q}_{\pmb{\theta },\pmb{\sigma }}[\mathcal{X}]]$ is the Gaussian approximation of the integrand evaluated at its maximum value.
Under the saddle-point approximation the posterior is:

\begin{align}
\label{eq:sppost}
\mathcal{P}[\pmb{\theta },\pmb{\sigma }|p[\mathcal{X}]] &\propto \mathcal{P}[\pmb{\theta },\pmb{\sigma }]\int_\mathcal{Q} e^{H[q[\mathcal{X}]]-\pmb{1}_D\cdot \left| \frac{\vect{g}_{\vect{\theta }}[\mathcal{X}]}{\pmb{\sigma }}\right| {}^2\cdot q[\mathcal{X}]}e^{-N H[p[\mathcal{X}]\|q[\mathcal{X}]]}dq[\mathcal{X}]\\
\label{eq:sppost2}
&\approx \mathcal{P}[\pmb{\theta},\pmb{\sigma }]e^{-N H\left[p[\mathcal{X}]\left\|\hat{q}_{\pmb{\theta },\pmb{\sigma }}[\mathcal{X}]\right.\right]}
\end{align}

This expression will be a good approximation to the exact model posterior when the system is close to statistical equilibrium. When the system is not close to statistical equilibrium the integral in \ref{eq:sp-int} will not be sharply peaked and saddle-point integration will not provide be a good approximation to the model posterior distribution. In these circumstances we would have to evaluate \ref{eq:contentprior2} explicitly.

\section{Examples}

We consider three examples that illustrate distinct informational settings. The first considers the scenario when the only information is in the form or theoretical or experimental information that determined an expected value that constrains the hypothetical frequency distributions describing the system. In this example the posterior distribution is proportional to the prior. The second example considers linear regression, the scenario when the information is in the form of direct observations that partially reveal the state of the system as well as in a specification of a linear relationship between observable variables with an approximate residual constraint.  In this example the posterior distribution is proportional to the Gaussian likelihood when a uniform prior over the model parameters is prescribed. In the third example, we incorporate both forms of information as a mixture of moment constraints and direct sample observations. 

We first consider the two extreme examples. In the simplest setting when there is no data sample to fit ($N=0$) and there are no theoretical constraints the posterior reduces to the entropy-favoring prior alone:

\begin{align}
\mathcal{P}[q[\mathcal{X}]|p[\mathcal{X}]]\propto e^{H[q[\mathcal{X}]]}
\end{align}

Maximizing the posterior implies a uniform distribution over the $K$ outcomes of the system, and puts all the prior probability on the uniform distribution at the center of the simplex $\mathcal{Q}_K$. While the situation of no information is of limited scientific interest it helps to illustrate the type of inferential reasoning associated with games of chance and is consistent with Laplace's principle of insufficient reason. 

In the case where there is data in the form of relative frequencies $p[\mathcal{X}]$ but no theoretical or experimental information in form of moment constraints the posterior reduces to the entropy-favoring prior and the multinomial likelihood:

\begin{align}
\mathcal{P}[q[\mathcal{X}]|p[\mathcal{X}]]\propto e^{H[q[\mathcal{X}]]}e^{-N H[p[\mathcal{X}]\|q[\mathcal{X}]]}
\end{align}

Maximizing the posterior will compromise between maximizing entropy in the prior and fitting the data in the likelihood. This situation is more often encountered in social sciences where observed data samples are the only information incorporated into the statistical analysis. 

We now consider an example from statistical physics to illustrate the situation of theoretical constraints with no data and approximate constraints on the data with a linear hypothesis, the example of linear regression, and an example of a mixture of moment and data constraints in a problem of statistical equilibrium in economic interactions. 

\subsection{Moment constraints in statistical mechanics}

An elementary problem in statistical physics is describing the state of an ideal gas, closed off from external influences, that consists of a very large number, $N$, of identical rapidly moving particles undergoing constant collisions with each other and the walls of their container. When $N$ is large (on the order of Avogadro’s number) it becomes informationally infeasible to derive the thermodynamic features of the gas, such as the temperature and pressure of the gas from a complete microscropic description of the gas, which would require $3 N$ coupled second order differential equations describing the time evolution of the positions of the particles in three-dimensional space. Information in the form of direct observation of the state of the system the momentum and position of each molecule is not feasible. As Maxwell and Boltzmann showed, however, we can create an ensemble of systems with the same coarse-grained state space each described by the same set of microscopic forces and macroscopic thermodynamic variables. It is then possible to draw powerful conclusions about ensembles despite our inability to solve the equations of motion for any one of the constituent systems. In this situation, a model consists of a frequency distribution over the positions and momenta of the particles constituting the system and the average values observed for the system constitute constraints the models must meet. For example, if the total energy is specified, and since the energy, $E[.]:\mathcal{X}\rightarrow \mathbb{R}_{\ge 0}$, of each particle depends on its momentum, the hypothesis of a given energy constitutes a constraint on the average energy across all particles:

\begin{align}
\mathcal{E}_{q[.]}[E[x]] &= \sum_\mathcal{X} E[x]q[x] = \bar{E}
\label{eq:energyconst}
\end{align}

If we know $\bar{E}$ we can eliminate any model that is inconsistent with the expected value constraints since they are impossible as explanations of the behavior of the system. Because the number of constraints imposed by the observed averages will most likely be smaller than the degrees of freedom allowed for the parameters characterizing the model the problem is underdetermined. We deal with the indeterminacy by assigning a prior probability that ranks the possible models consistent with the constraints.

\begin{align}
\begin{aligned}
&\max_{q[\mathcal{X}]\in \mathcal{Q}_K} -\sum_\mathcal{X}q[x]\log[q[x]] \\
&\text{subject to } \sum_\mathcal{X} E[x]q[x] = \bar{E} 
\end{aligned}
\end{align}

We construct the Legendre transform by maximizing entropy subject to the model constraints, which results in a frequency distribution parameterized by the Lagrange multipliers of the CME problem:

\begin{align}
\begin{aligned}
\hat{q}_{\lambda }[\mathcal{X}]\propto e^{-\lambda E[x]} = e^{-\frac{E[x]}{\sigma^2}}
\end{aligned}
\end{align}

Because in this case there is no likelihood component to Bayes’ Theorem the posterior probability distribution \ref{eq:post} is just proportional to the constrained prior probability distribution and entropy favoring component of the prior:

\begin{align}
\label{eq:postex}
\mathcal{P}[q[\mathcal{X}]|p[\mathcal{X}]]\propto e^{H[q[\mathcal{X}]]}e^{-\frac{E[x]}{\sigma^2}}=e^{-H[q[\mathcal{X}]||\hat{q}_\sigma[\mathcal{X}]]}
\end{align}

The posterior distribution assigns probabilities to distributions in the simplex $\mathcal{Q}_K$ proportional to the exponential of the Kullback-Leibler divergence of the hypothesis distribution to the CME distribution. When the constraint tolerance $\sigma>0$ the posterior assigns a positive probability to every distribution in the simplex $\mathcal{Q}_K$ consistent with the information. The prior probability effectively ranks the possible models consistent with the constraints thus dealing with indeterminacy arising from the underdetermination of the problem.

\subsection{Sample observation constraints in linear regression}

In linear regression the system state typically consists of $R$ observed variables that partially reveal the state of the system and interest lies in the correlation between the $R-1$ variables that constitute the ``design matrix" of independent variables and the $R^{th}$ variable that defines the ``dependent variable".  The system state can also be specified as $R+1$ variables which includes the constant $1$ as the first variable so that $\vect{x} = \{1,x_1,\ldots,x_R\}$. The linear hypothesis is \[x_R = \beta_0 + \beta_1 x_1 + \ldots \beta_{R-1}x_{R-1}\] or $\vect{x}\cdot \{\vect{\beta},-1\}=0$ where $\vect{\beta}$ is an $R$-dimensional vector of regression coefficients. In this case there is a single unobserved vector of residuals that are the difference between the predicted and observed dependent variable, $g_\vect{\beta}[\vect{x}] = \vect{x}\cdot \{\vect{\beta},-1\}$. A model for the linear hypothesis makes assertions about the joint distribution of residuals. In this linear case, the imposition of exact constraints, $g_\vect{\beta}[\vect{x}]=0$, constraints the posterior distribution over the hypothesis to only those frequency distributions that assign positive probabilities to states that fall exactly on the hypothesized linear relationship. Since the likelihood of any real system satisfying the exact linear theoretical constraints is effectively zero, the system becomes overdetermined. If the constraints are allowed to hold approximately to some degree of tolerance we can write

\begin{align}
\label{eq:approxconstraints}
\sum_\mathcal{X}|\vect{g}_\vect{\beta}[x]|^2 q[x]= \left|\vect{g}_\vect{\beta}[\mathcal{X}]\right|^2 \cdot q[\mathcal{X}] =\vect{s}^2 
\end{align} 

The imposition of approximate constraints transforms an overdetermined problem into an underdetermined problem by introducing additional degrees of freedom in the vector of tolerances $\vect{s}^2$. Because the system is underdetermined the observable states of the system will now be compatible with more than one member of the class of models and an assumption of a prior probability is necessary to resolve this indeterminacy. The approximate constraint implies the scaled mean squared residuals $|\frac{\vect{g}_\vect{\theta}[\mathcal{X}]}{\vect{\beta}}|^2$ will form a $D \times K$ dimensional matrix, and we can write the constraint-favoring component of the prior:

\begin{align}
\label{eq:resprior}
\hat{q}_{\vect{\beta},\pmb{\sigma }}[\mathcal{X}]=e^{-\vect{1}_D\cdot \left|\frac{\vect{g}_\vect{\beta}[\mathcal{X}]}{\vect{\sigma}}\right|^2\cdot q[\mathcal{X}]}
\end{align}

If in general we favor hypothetical distributions that satisfy the constraints and also have higher entropy

\begin{align}
\label{eq:resentprior}
\mathcal{P}[q[\mathcal{X}]]\propto e^{H[q[\mathcal{X}]]}e^{-\vect{1}_D\cdot \left|\frac{\vect{g}_\vect{\beta}[\mathcal{X}]}{\vect{\sigma}}\right|^2\cdot q[\mathcal{X}]}=e^{-H\left[q[\mathcal{X}]\left\|\hat{q}_{\pmb{\beta},\pmb{\sigma }}[\mathcal{X}]\right.\right]}
\end{align}

As $\vect{\sigma}\rightarrow \infty$ the constraint \ref{eq:approxconstraints} becomes non-binding and the prior becomes proportional to $e^{H[q[\mathcal{X}]]}$. If the vector of tolerances are known, but the parameters of the linear model hypothesis are unknown then posterior estimation is done through evaluation of 

\begin{align}
\label{eq:posterior2}
\mathcal{P}[\pmb{\beta },q[\mathcal{X}]|p[\mathcal{X}],\pmb{\sigma }]\propto e^{H[q[\mathcal{X}]]-\pmb{1}_D\cdot \left| \frac{\vect{g}_{\vect{\beta }}[\mathcal{X}]}{\pmb{\sigma }}\right| {}^2\cdot q[\mathcal{X}]}e^{-N H[p[\mathcal{X}]\|q[\mathcal{X}]]}
\end{align}

We know from \ref{eq:sppost2} that if the system is close to statistical equilibrium (which is presumably what makes the hypothesis valid), the posterior is approximately proportional to:

\begin{align}
\mathcal{P}[\vect{\beta}|p[\mathcal{X}],\pmb{\sigma }] \approxprop \mathcal{P}[\vect{\beta}]e^{-N H\left[p[\mathcal{X}]\left\|\hat{q}_{\vect{\beta},\pmb{\sigma }}[\mathcal{X}]\right.\right]}
\end{align}
If both model parameters and tolerances are unknown we need to evaluate

\begin{align}
\mathcal{P}[\vect{\beta},\pmb{\sigma }|p[\mathcal{X}]] \approxprop \mathcal{P}[\vect{\beta},\pmb{\sigma }]e^{-N H\left[p[\mathcal{X}]\left\|\hat{q}_{\vect{\beta},\pmb{\sigma }}[\mathcal{X}]\right.\right]}
\end{align}

If we assign a uniform prior probability over the model parameters of the form $\mathcal{P}[\vect{\beta},\pmb{\sigma }]=\frac{d \vect{\beta}}{\pmb{\sigma }}$ the posterior becomes proportional to the likelihood. 

\subsection{Mixture of information: quantal responses in social interactions}

A third example that demonstrates how a mixture of moment constraints and information in the form of direct observations shapes posterior inference is the economic model of quantal response statistical equilibrium (QRSE) developed in \citet{SchFoley2017, Scharfenaker2020a, Scharfenaker2020, ScharfenakerFoley2023, SchPds2019}. In this case we have observable data on a social outcome that is brought into a statistical equilibrium by the purposive behavior of individual agents interacting in a decentralized competitive environment. For example, Adam Smith's theory of profit rate equalization is based on the theory that the disposition of competitive capitalist firms implies that capital will move away from sectors of production with low profit rates and crowd into sectors with high profit rates. The movement of capital has the unintended consequence of lowering the profit rate in those sectors with above average profit rates, and raising profit rates for sectors with below average profit rates. The negative feedback of the mobility of capital generates strong statistical regularities in the distribution or profit rates \citep{Scharfenaker2017,Farjoun1983} that reflects an underlying statistical equilibrium in firm states. In this scenario we have access to information in form of data from firms' balance sheets that can be used to calculate their rate of profit, as well as information in form of theoretical constraints on the average behavior of capitalist firms and the negative feedback of capital investment on the rate of profit.   

The individual components of the system are competitive capitalist firms. At any moment in time each firm characterized by its profit rate state, $x$ representing the social outcome, and a quantal action variable $a\in \{\text{entry, exit}\}$, representing a behavioral theory of investment of capitalist firms who either entering or exiting the sector of production in which its profit rate is observed. The theoretical state of the system is a joint frequency distribution over profit rates and quantal actions $f[x,a]$. 

Theoretical considerations on profit rate equalization imply informational constraints on the conditional frequency $f[a|x]$, representing the tendency of a typical firm to exit markets with low profit rates and enter markets with high profit rates, and the conditional frequency $f[x|a]$, which represents the negative impact of firm entry on profit rates and positive impact of firm exit on profit rates. If we constrain firm behavior to be probabilistically described by the softmax function that reflects the decentralized nature of competition we have:

\begin{align}
\label{eq:qrprentry}
f[\mbox{entry}|x] =& \frac{1}{1+e^{-\frac{x-\mu}{T}}}\\
f[\mbox{exit}|x] =& \frac{1}{1+e^{\frac{x-\mu}{T}}}
\end{align}

\noindent where $\mu$ is the profit rate at which firms are indifferent between entering and exiting and $T$ is the scale of behavioral fluctuations representing the sensitivity of firms to profit rate differentials. 

The constraint representing the theoretical information on the negative feedback of entry/exit decisions on profit rates can be expressed in terms of a limiting difference between the expected profit rates conditional on entry and exit.

\begin{align}
\label{eq:compconst}
f[\mbox{entry}]\mathcal{E}_f[x-\alpha|\mbox{entry}] - f[\mbox{exit}]\mathcal{E}_f[x-\alpha|\mbox{exit}]\le \delta
\end{align}

The parameter $\alpha$ represents the central tendency toward which competition pushes the profit rate in response to entry and exit. This is a constraint on conditional expectations parallel to the constraints in statistical mechanical models. The parameter $\delta$ that represents the limits to the differences in conditional expectations is inherent in the Smithian theory of competition. Even though $\delta$ cannot be observed directly from data it is accessible indirectly through the methods of Bayesian inference, since it is part of the hypothesis in the Bayesian framework.\footnote{See \citet{Golan1996} for similar application using maximum entropy.} From the definition of conditional expectations we can write the constraint \ref{eq:compconst} in terms of the conditional frequencies $f[a|x]$ and use the specification in \ref{eq:qrprentry} to simplify the expression of \ref{eq:compconst} as: 

\begin{align}
\label{eq:compconst2}
\int_{\mathcal{X}} \tanh \left[\frac{x-\mu}{T}\right]f[x](x-\alpha) dx \le \delta
\end{align}

Maximizing the entropy of the joint distribution $f[x,a]$ representing the state of the system subject to the theoretical constraints of profit rate equalization defines the Constrained Maximum Entropy problem\footnote{Where we use the decomposition of joint entropy into the marginal and conditional entropies to solve the problem.}:

\begin{align}
\label{eq:compCME}
\max_{f[x,a]}\, & H[f[x,a]]=H[f[x]]+ \int_{\mathcal{X}} f[x]H[f[a|x]] dx\\
\label{eq:compCME1}
\text{subject to }&\int_{\mathcal{X}} f[x] dx = 1\\
\label{eq:compCME2}
\text{and }&\int_{\mathcal{X}} \tanh \left[\frac{x-\mu}{T}\right]f[x](x-\alpha) dx = \delta
\end{align}

The CME problem can be solved for the marginal frequencies of observable profit rates:

\begin{align}
\label{eq:QRSE}
\hat{f}[x] \propto& e^{H_{\mu,T}[f[a|x]]}e^{-\tanh \left[\frac{x-\mu}{T}\right]\frac{x-\alpha}{S}}
\end{align}

The parameter $S$ is the inverse of the Lagrange multiplier corresponding to the competition constraint (\ref{eq:compconst}). It has the same dimensions as the profit rate, and can be interpreted as the scale on which competition equalizes the profit rate. The posterior distribution is described for the joint hypothesis over $q[\mathcal{X}],\mu,T,\alpha,S$. If profit rate equalization is achieved on a time scale short enough for us to treat the observations as coming from a system in statistical equilibrium the posterior will be approximately equal to 

\begin{align}
\begin{aligned}
\label{eq:QRSEposterior}
\mathcal{P}^{QRSE}[\mu,T,\alpha,S|x] \propto \mathcal{P}[\mu,T,\alpha,S] e^{H_{\mu,T}[f[a|x]]}e^{-\tanh \left[\frac{x-\mu}{T}\right]\frac{x-\alpha}{S}}
\end{aligned}
\end{align}

With an appropriate coarse graining of profit rates, the system can be described in terms of the multinomial likelihood \ref{eq:multlike}.

\begin{align}
\mathcal{P}[\mu,T,\alpha,S|p[\mathcal{X}]] \approxprop \mathcal{P}[\mu,T,\alpha,S]e^{-N H\left[p[\mathcal{X}]\left\|\hat{q}_{\mu,T,\alpha,S}[\mathcal{X}]\right.\right]}
\end{align}

where 

\begin{align}
\hat{q}_{\mu,T,\alpha,S}[\mathcal{X}]\propto& e^{H_{\mu,T}[f[a|x]]}e^{-\tanh \left[\frac{x-\mu}{T}\right]\frac{x-\alpha}{S}}
\end{align}

In the Bayesian framework the solution to the CME problem \ref{eq:QRSE} provides the likelihood for inferences concerning the parameters $\mu,T,\alpha,S$ given appropriate priors. The information corresponding to the theoretical constraints implied by profit rate equalization determines the class of distributions consistent with the information. This information shapes the posterior distribution by modifying the prior in \ref{eq:post}. When the system is close to statistical equilibrium and the Lagrange multipliers are unknown, they become part of the hypothesis and posterior inference centers on estimating their values conditional on observed profit rates. In this example we have a mixture of information, the theoretical moment constraint \ref{eq:compconst2} and profit rate data from firm balance sheets. The theoretical information determines the class of distributions \ref{eq:QRSE} consistent with the information that imply a likelihood for Bayesian estimation given information in the form of direct observations of firm profit rates.

\section{Conclusion}

Bayes' theorem tells us that the joint assignment of probabilities as degrees of belief to combinations of data realizations and hypotheses is a fundamental building block of statistical analyses. Any consistent and general framework for assigning posterior degrees of belief must be able to incorporate information in the form of a mixture of moment constraints and direct observations that partially reveal the state of the system. In situations where we have information in the form of a mixture of theoretical or experimental constraints on the hypothesis as well as a direct observations of the state of system, we can adopt a constrained entropy-favoring prior to reflect the degree to which the hypothesis meets the constraints and then apply the observations through a likelihood function to the prior in order to compute the posterior conditional on the available information. 

For general multinomial systems the posterior distribution is a joint assignment of frequencies $\mathcal{P}[q[\mathcal{X}],p[\mathcal{X}]]$ that implies a likelihood and prior $\mathcal{P}[p[\mathcal{X}]|q[\mathcal{X}]]\mathcal{P}[q[\mathcal{X}]]$. Information in the form of direct observations of the state of the system determine the relative frequencies of states $p[\mathcal{X}]$ requiring a reassignment of probabilities in the posterior distribution to be consistent with the data observations. When information rules out certain configurations of the state space the principle of maximum entropy incorporates this information in the form of constraints that make the prior distribution consistent with the information parameterized in term of the Legendre transform $\hat{q}_{\vect{\theta},\vect{\sigma}}[\mathcal{X}]$, where $\vect{\theta},\vect{\sigma}$ represent the model parameters and tolerances of approximate constraints. If the constraints are known exactly or approximately posterior inference follows from directly evaluating the conditional probability $\mathcal{P}[q[\mathcal{X}],p[\mathcal{X}]|\vect{\theta},\vect{\sigma}]$. When constraint values are unknown they become part of the hypothesis and posterior inference follows from the joint evaluation of $\mathcal{P}[q[\mathcal{X}],p[\mathcal{X}],\vect{\theta},\vect{\sigma}]$. When interest centers on the value of the model parameters marginalization over the hypothesis distribution $q[\mathcal{X}]$ and the system is close to statistical equilibrium the posterior can be approximated by the entropy maximizing likelihood and prior over model parameters. In this situation the relationship between the principle of maximum entropy and Bayes' theorem is clear.

\section*{Acknowledgments}

We would like to thank Amos Golan for many useful discussions on these topics.

\bibliography{BayesMaxEnt.bib}

\end{document}